\documentclass[a4paper]{article}
\pdfoutput=1
\usepackage[pages=all, color=black, position={current page.south}, placement=bottom, scale=1, opacity=1, vshift=5mm]{background}
\SetBgContents{
	Source code available: \url{https://github.com/xwasco/DominantSetLibrary}
}      

\usepackage[margin=1in]{geometry} 

\usepackage{amsmath}
\usepackage{amsthm}
\usepackage{amssymb}

\usepackage[utf8]{inputenc}
\usepackage{hyperref}
\hypersetup{
	unicode,
	pdfauthor={Sebastiano Vascon, Samuel Rota Bul`o, Vittorio Murino, Marcello Pelillo},
	pdftitle={DSLib:  An open source library for the dominant set clusteringmethod},
	pdfsubject={DSLib:  An open source library for the dominant set clusteringmethod},
	pdfkeywords={Dominant set, clustering, graph, maximal clique, game theory},
	pdfcreator={pdflatex}
}

\usepackage[sort&compress,numbers,square]{natbib}
\theoremstyle{plain}

\theoremstyle{definition}

\usepackage{graphicx, color}
\graphicspath{{fig/}}

\usepackage[framed,numbered,autolinebreaks,useliterate]{mcode}
\usepackage{mathrsfs} 

\newcommand{\matlab}{Matlab\textsuperscript{\textregistered} }

\usepackage{lipsum}

\title{DSLib: An open source library for the dominant set clustering method}
\author{
Sebastiano Vascon\\
Ca' Foscari University of Venice\\
sebastiano.vascon@unive.it
\and
       Samuel Rota Bulò\\
       Mapillary Research\\
       samuel@mapillary.com
       \and
       Vittorio Murino\\
       Istituto Italiano di Tecnologia\\
       vittorio.murino@iit.it \\
        \and
       Marcello Pelillo\\
       Ca' Foscari University of Venice\\
pelillo@unive.it}

\begin{document}
	\maketitle
	
	\begin{abstract}
DSLib is an open source implementation of the Dominant Set (DS) clustering algorithm written entirely in \matlab. The DS method is a graph-based clustering technique rooted in the evolutionary game theory that starts gaining lots of interests in the computer science community. It has been originally introduced in \citep{Pavan2003, Pavan2007} and, thanks to its duality with game theory and its strict relation to the notion of maximal clique, has been explored in several directions not only related to clustering problems. Applications in graph matching, segmentation, classification and medical imaging are common in literature. 
This package provides an implementation of the original DS clustering algorithm, since no code has been officially released yet, together with a still growing collection of methods and variants related to it.
Our library is integrable into a \matlab pipeline without dependencies, it is simple to use and easily extendable for upcoming works.
The latest source code, the documentation and some examples can be downloaded from \url{https://xwasco.github.io/DominantSetLibrary}
		
	\end{abstract}
	
\section{Introduction}
 
Clustering is a relevant task in the domain of pattern recognition.
 
It requires to group
elements into meaningful subsets (e.g., points which are close in a Euclidean space) exploiting the underlying hidden structure of the data. 

It is in general an ill-posed problem
since different partitioning of the same dataset could end up into a meaningful subdivision of the data. 
Two properties characterize, in general, every clustering algorithms: each subset (cluster) should have a \emph{high internal coherence} and a \emph{high external in-coherence} this means that items belonging to the same cluster should have a high similarity while items of different clusters should be more dissimilar as possible. These properties are essential to achieving a good separation of the subsets and represent the foundation of many clustering algorithms 

including the dominant set clustering method.
 The dominant set (DS) algorithm has been formalized in the domain of graph theory, with a strict relationship to the notion of maximal clique \footnote{A clique is a set of mutually connected nodes} generalizing it to the edge-weighted case.

In the DS algorithm, a dataset is represented as an edge-weighted graph where nodes are data points, and edges encode the relationship between pairs of points in terms of a similarity weight.

The graph, encoded as an affinity matrix, is given in input to the algorithm.
The clusters (dominant set) are then iteratively extracted from the graph using a dynamical system that mimics a natural selection process. Three different dynamical systems are implemented into this library.

In particular, using this clustering technique is beneficial when \emph{i)} the number of clusters to be found is totally unknown, \emph{ii)} the pairwise similarity function is not symmetric, \emph{iii)} a quantification of the quality of cluster is needed (i.e. to detect outliers) and \emph{iv)} a notion of centrality of an element with respect to the cluster (centroid) is required. The aforementioned properties are hard to find in a traditional clustering algorithm (i.e. k-means, spectral clustering and so on). Furthermore, the DS algorithm proved to be particularly suitable when the datasets are noisy and corrupted by outliers thanks to its nature of finding highly compact and coherent structures in the graph.\\
~\\
The remaining of the paper is organized as follow: in Sec \ref{sec:ds} a brief introduction to the dominant set method and evolutionary game theory are provided, in Sec \ref{sec:sp} the software package is explained with a practical example, in Sec \ref{sec:case} some use-cases are given and, finally, in Sec \ref{sec:conclusion} conclusions are drawn.

\subsection{Dominant Sets}\label{sec:ds}
A dataset is represented as an edge-weighted graph $G=(V, E, \omega)$ with no self-loops, where $V$ is the set of nodes, $E$ is the set of edges connecting pairs of nodes and $\omega:E\to\mathbb R_+$ is a weighting function mapping edges to nonnegative weights. One of the possible way of defining $\omega$ is through the following kernel $\omega(i,j)=e^{-\frac{d(i,j)}{\sigma}}$ where $d(i , j)$ is a generic distance function (typically the Euclidean distance) between elements $i$ and $j$ (with $i \neq j$) and $\sigma$ is the decay of the negative exponential (or the scale of the kernel). Since the DS method requires a graph in terms of its pair-wise affinity matrix $A$, other similarity functions can be plugged in by the user without any problem.\\
Finding a dominant set (a cluster) is achieved optimizing a quadratic problem over the standard simplex:

\begin{equation}\label{eqn:program}
\begin{aligned}
\max&\qquad\textbf{x}^\top A\textbf{x}\\
\mbox{s.t. }&\qquad\textbf{x} \in \Delta^n
\end{aligned}
\end{equation}
where $A$ is the affinity matrix of size $n \times n$, $n=|V|$ the number of nodes in the graph (number of elements in the dataset) and $\Delta^n$ is the standard simplex consisting of $n$-dimensional probability vectors.
A local solution of problem \eqref{eqn:program} can be found by iterating the following dynamics until convergence:
\begin{equation}
		x_i(t+1)=x_i(t)\frac{(A\textbf{x}(t))_i}{\textbf{x}(t)^TA\textbf{x}(t)} \label{eqn:repdyn}
\end{equation}
Eq.\eqref{eqn:repdyn} is a dynamical system, known as \emph{Replicator Dynamics}(RD) \citep{weibull1995}, which mimics a Darwinian selection process over the nodes, letting the ones that mostly support each other (the ones that are more similar) to survive (retain a nonzero probability) to the detriment of the others. Two additional dynamical systems are proposed in this package, the Infection Immunization \citep{Rota2011} and the Exponential Replicator Dynamics \citep{weibull1995} which can be used for large graphs and for faster convergence, respectively. 
At convergence of the chosen dynamical system, the elements that exceed a given probability threshold $\theta$ represent the dominant set. Those elements are removed from the original graph and the process is reiterated on the remaining nodes until all the nodes are grouped. This iterative peeling-off procedure to extract DSs is implemented in the library (see Sec \ref{sec:practical}).

\section{Description of the package}\label{sec:sp}
In this section, we provide a short description of the library and how to integrate it into a \matlab pipeline. The inline documentation of the package provides all the information for the parameters.

\subsection{Requisites}
No particular requirements are needed, the library runs on any Matlab environment (Windows, Linux, Mac) and it has no dependencies. It has been tested on Matlab 2014b, 2016b and 2018b in Windows/Linux and MacOSX. Some components are written in C-mex for optimization purpose but equivalent Matlab code is available without the need of compiling it.

\subsection{What's inside the package?}
In this version of the library, we have implemented the following papers: \citep{Pavan2003, Pavan2007, Rota2011, Vascon2013} and a still growing set of papers will be added in the future. For the curious reader, we refer to \citep{bulo2017} for an outstanding review of the DS method.

\subsection{Practical usage}\label{sec:practical}
In this section we show how to setup a basic running example to cluster a synthetic dataset.\\
\textbf{Step 1: Generating data and build the affinity matrix $A$}:

\begin{lstlisting}
rng('default'); 							% For repeatability
cx = [1 1;5 5 ;8 8];					%center of the clouds of points 
npts=100; pts= repmat(cx,npts,1) + randn(npts*size(cx,1),2);
A=pdist(pts); 							%pairwise Euclidean distances
s=3*var(A);		%a heuristic to find sigma 
A=exp(-A./s);  A=squareform(A);	%from distance to similarity
A=A.*not(eye(size(A))); %the graph should not have self-loops
\end{lstlisting}

\textbf{Step 2: Choosing the evolutionary dynamic and the DS parameters}:

\begin{lstlisting}
dynType=1;  		%0=Replicator Dynamics, 1=InfectionImmunization 2=Exponential replicator dynamics
precision=1e-6;  %the precision required from the dynamical system
maxIters=1000; 	%number of maximum iteration of the dynamical system
x=ones(size(A,1))./size(A,1); %starting point of the dynamical system
theta=1e-5;		%threshold used to extract the support from x.
\end{lstlisting}
\textbf{Step 3: Call the clustering method}

\begin{lstlisting}
[C]=dominantset(A,x,theta,precision,maxIters,dynType);
\end{lstlisting}
alternatively, the clustering method can be called providing only the similarity matrix and using the default parameter.

\begin{lstlisting}
[C]=dominantset(A);
\end{lstlisting}
\textbf{Step 4: Show the cluster results}

\begin{lstlisting}
gscatter(pts(:,1),pts(:,2),C);	%show the points colored by cluster
\end{lstlisting}
The package comes with more complex examples and a deep inline documentation.

\section{Case studies}\label{sec:case}
The core described in this manuscript has been used in several studies. Here we highlight three applications in radically different domains to prove the transversality of the method:
\begin{itemize}
\item \textbf{Protein Clustering}: In \citep{Pennacchietti2017} the DS has been used to study the interaction of proteins during the neuronal plasticity. The tool provides unprecedented insight into the problem. Number of nodes in the graph $\simeq 10.000$.

\item \textbf{Conversational group detection}: In \citep{Vascon2014,Vascon2016} the DS has been used to automatically detect conversational groups in real-time in video sequences, a task of utmost importance in video surveillance. Number of nodes in the graph $\simeq 30$.

\item \textbf{Brain connectomics}: In \citep{Dodero2013,Dodero2015}  the DS has been used to study the neuronal connections between the hemisphere with the aims of automatically group fibers having the same shape.   Furthermore,  a matching procedure allows recovering the common structures across different subjects. Number of nodes in the graph $\simeq 30.000$.
\end{itemize}

\section{Conclusion}\label{sec:conclusion}
The DS clustering method has proven its powerfulness in the last decade,  leading to a variety of applications.   Here  we  present  a  library  implementing  the  original paper  and  some  follow-up  methods.  The library is written in  pure  \matlab with C-mex components, is cross-platform, open source and does not have any dependency.  Full documentation, tutorials, download,  and  the  updated  source  code  can  be  found  at \url{https://xwasco.github.io/DominantSetLibrary/}.
\section*{Acknowledgments}
We would like to acknowledge Dr. Luca Dodero, prof. Diego Sona for testing the library and giving us their invaluable support. Furthermore, we want to thanks the entire \emph{Student Branch @ DAIS} of Ca' Foscari University of Venice for their suggestions.
	
\bibliography{main}

\end{document}